\pdfoutput=1
\documentclass[twocolumn,aps,prb,10pt,floatfix,superscriptaddress]{revtex4-2}
\newcommand{\up}{\uparrow}
\newcommand{\down}{\downarrow}

\usepackage{graphicx}
\usepackage{bm}
\usepackage{dcolumn}
\usepackage{amssymb}
\usepackage{amsmath}
\usepackage{amsbsy}
\usepackage{amsfonts}
\usepackage{epsfig}
\usepackage{graphicx}
\usepackage{multirow}
\usepackage{stackrel}
\usepackage{paralist}
\usepackage[rgb]{xcolor}
\usepackage[textsize=tiny,textwidth = 1.6 cm]{todonotes}

\usepackage[percent]{overpic}
\usepackage{hyperref}
\usepackage{tikz}
\usetikzlibrary{intersections} 
\usetikzlibrary{positioning,calc}
\usepackage[caption=false]{subfig}
\usepackage[normalem]{ulem}

\tikzset{
    right angle quadrant/.code={
        \pgfmathsetmacro\quadranta{{1,1,-1,-1}[#1-1]}     
        \pgfmathsetmacro\quadrantb{{1,-1,-1,1}[#1-1]}},
    right angle quadrant=1, 
    right angle length/.code={\def\rightanglelength{#1}},   
    right angle length= 1 ex, 
    right angle symbol/.style n args={3}{
        insert path={
            let \p0 = ($(#1)!(#3)!(#2)$) in     
                let \p1 =  ($(\p0)!\quadranta*\rightanglelength!(#3)$),
                \p2 = ($(\p0)!\quadranta*\rightanglelength!(#2)$) in 
                let \p3 = ($(\p1)+(\p2)-(\p0)$) in  
            (\p1) -- (\p3) -- (\p2)
        }
    }
}

\newcommand{\cancel}[1]{}

\definecolor{NewColor}{rgb}{1,0,0}
\definecolor{myRed}{rgb}{1,0,0}
\definecolor{myGreen}{rgb}{0.2,0.6,0.2}
\definecolor{myBlue}{rgb}{0,0,1}

\graphicspath{{../Graphics/}{../Graphics/Publication/}}

\newcommand{\CO}[1]{\textcolor{red}{}}

\newlength{\noLengthl}

\renewcommand{\vec}[1]{{\boldsymbol{#1}}}

\begin{document}
\title{
Large diversity of magnetic phases in two-dimensional magnets with spin-orbit coupling and superconductivity
}

\author{Jannis Neuhaus-Steinmetz}
\affiliation{Department of Physics, University of Hamburg, 20355 Hamburg, Germany}
\author{Tim Matthies}
\affiliation{Department of Physics, University of Hamburg, 20355 Hamburg, Germany}
\author{Elena Y. Vedmedenko}
\affiliation{Department of Physics, University of Hamburg, 20355 Hamburg, Germany}
\author{Thore Posske}
\affiliation{I. Institute for Theoretical Physics, University of Hamburg, D-20355 Hamburg, Germany, The Hamburg Centre for Ultrafast Imaging, Luruper Chaussee 149, 22761 Hamburg, Germany.}
\author{Roland Wiesendanger}
\affiliation{Department of Physics, University of Hamburg, 20355 Hamburg, Germany}

\begin{abstract}
We classify the magnetic ground states of a 2D lattice of localized magnetic moments which are coupled to a superconducting substrate with Rashba-spin-orbit coupling. We discover a rich magnetic phase diagram with surprisingly complex structures including 2q-spin-spirals, a 2x2-periodic pattern, and skyrmion lattices, self-consistently, using an effective classical spin Hamiltonian and show that the system hosts non-zero $4$-spin interactions.
Our in-depth analysis of about ten thousand magnetic configurations becomes feasible using contrastive clustering, a recent advanced unsupervised machine learning technique.
This work proposes simple few-band systems for non-collinear magnetic states and stimulates further research on topological effects in their self-consistent electronic structure.
\end{abstract}

\maketitle

\section{Introduction}
There is a vast versatility of magnetic phases in the physics of interfaces.
Among ferromagnetic and antiferromagnetic spin orientations in the ground state, recently noncollinear phases including spirals, multi-q states \cite{2q, Heinze2011, Bedow2020_3Q}, canted AFMs \cite{Canted}, bubbles, and topological magnetism including skyrmions \cite{doi:10.1126/science.1240573} and merons \cite{Shao2023}, as well as further exotic phases like altermagnets \cite{PhysRevX.12.040501} have been theoretically and experimentally discovered. Magnetic phases form a cornerstone of today's and future storage technology \cite{Fert2013, PhysRevLett.57.2442, Back_2020} and are fundamentally interesting because of exotic classical and quantum quasiparticle excitations that could be hosted in frustrated or correlated magnets \cite{PhysRevLett.116.077202,Kitaev_2006}. For modeling purposes and for detecting yet novel magnetic phases, the question arises which conceptionally simple models host a large number of magnetic phases that can be tuned by experimentally accessible parameters. Particularly interesting in this regard are systems combining magnetism and superconductivity.
While numerous studies of one-dimensional magnetic systems in proximity to s-wave superconductors
exist, motivated by the interest in Majorana physics \cite{1D, Schneider2022, Feldman2017, new14}, the data on phase diagrams of their two-dimensional counterparts have been very limited up to now \cite{new15-1,new15-2}. Ground state studies on quantum and classical magnetic moments coupled to itinerant electrons found evidence of chiral order \cite{PhysRevLett.101.156402}, topological electronic order \cite{ido2024manybody, JPSJ.79.083711}, and multi-spin interactions \cite{PhysRevLett.108.096401} in frustrated triangular lattices without superconductivity. Motivated by recent studies on superconducting 1D systems \cite{1D}, we therefore consider a two-dimensional square lattice tight-binding model with local magnetic moments, Rashba spin-orbit coupling, and superconductivity. Our model is effectively describing localized classical magnetic moments that are coupled by the itinerant electrons of a surface state, not specifically bound to any parameter regime.
The knowledge of possible ground states in those complex structures  is required for several reasons. First, exotic topological states have been found in 2D itinerant magnets in proximity to a superconductor (SC) \cite{new15-1,new15-2}. Second, interesting multi-spin interactions can be present in such systems \cite{PhysRevB.77.064410, 2q} that might lead to unexpected magnetic and electronic phases.
Due to big datasets in the study of magnetic systems,  machine learning has been applied extensively in this field, which we adopt as well. Corresponding techniques have especially been employed to predict the parameters of the magnetic Hamiltonian \cite{kawaguchi2021determination}, to retrieve the topological charge from dynamic chiral magnetic structures \cite{matthies2022topological}, and to construct phase diagrams for skyrmionic systems \cite{albarracin2022machine,albarracin2024unsupervised}. Additionally, unsupervised learning has found a variety of applications in the field of microscopic magnetism \cite{kwon2022autoencoderGroundState, yoon2022interpolation, park2022optimization} and the identification of topological phases~\cite{kaming2021unsupervised,rem2019identifying}.

In this manuscript, we analyze the ground states of magnetic moments coupled by a spinful tight-binding model with proximity induced s-wave superconductivity, local magnetic Zeeman moments and Rashba-spin-orbit-coupling (RSO). The resulting approximately $10000$ magnetic configurations are too diverse to be reliably classified by a simple algorithm.
We therefore use contrastive learning~\cite{chen2020simple,caron2020unsupervised}, a recent unsupervised machine learning technique, which can be adapted to the physical symmetries. While contrastive learning was used to classify phases in the Ising, Compass, and Su-Schrieffer-Heeger models~\cite{Han_2023} using the schema of Ref.~\cite{chen2020simple}, we used Contrastive Clustering~\cite{li2021contrastive} to obtain the phases directly from the network. Our results show that this technique is highly effective for the classification of complex phase spaces of magnetic systems on SC.

This paper is structured as follows. In Sec.~\ref{model}, we introduce the tight-binding model coupled to classical spins, our method to calculate its magnetic ground states, and explain how we used an artificial neural network to classify the magnetic ground states. In Sec.~\ref{results}, we discuss the magnetic properties of the respective ground states. In Sec.~\ref{conclusions}, we summarize our results and give an outlook.

\section{Model and Method} \label{model}
\subsection{System}
\begin{figure}
    \centering
    \includegraphics[width=0.45\textwidth]{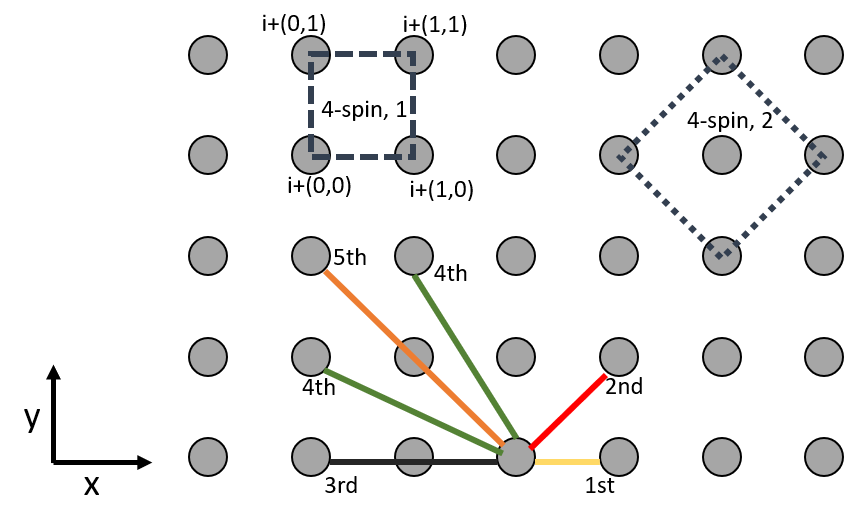}
    \caption{Spin interactions used in Eq.~\eqref{H_H}. The gray dots represent spins on atomic sites. The filled lines mark  $2$-spin interactions with the 1st (yellow), 2nd (red), 3rd (black), 4th (green), and 5th (orange) nearest neighbor. The dotted lines show the two considered $4$-spin interactions with \textit{1} and \textit{2} denoting the smaller and larger square, respectively.}
    \label{spin_interactions}
\end{figure}

We investigate a two-dimensional electron square lattice with classical local magnetic moments, proximity-induced s-wave superconductivity, Rashba spin-orbit coupling (RSO), and periodic boundary conditions. The system is described by the following Hamiltonian
\begin{align}\label{H_TB}
H= & \sum_{j=1}^N \vec{c}^{\dagger }_j\Big(-J^{tb}{\tau }_0\vec{m}_j\cdot \vec{\sigma }-B_z\tau_0\sigma_z \\
 &-\mu {\tau }_z{\sigma }_0+\Delta\tau_x\sigma_0\Big)\vec{c}_j\nonumber \\
+\sum_{<i,j>_1}&{\vec{c}^{\dagger }_i\Big(t{\tau }_z{\sigma }_0+i\alpha\left[d_x(i,j)\tau_z\sigma_y-d_y(i,j)\tau_z\sigma_x \right]\Big)\vec{c}_j}  \nonumber
\end{align}
with the Nambu spinor $\vec{c}_j=(c_{j\up },c_{j\down },c^{\dagger }_{j\down },{-c}^{\dagger }_{j\up })$ \cite{nambu}, the coupling $J^{tb}$ between a magnetic moment on a given site $j$ and the spin of an electron, the orientation of the local magnetic moments on the $j$-th site $\vec{m}_j$, the chemical potential $\mu$, the hopping amplitude $t$, the superconducting order parameter $\mathrm{\Delta }$, the strength of a magnetic field perpendicular to the lattice $B_z$, and the strength of the Rashba-spin-orbit coupling $\alpha$. The difference in $x$ and $y$ position of the sites $i$ and $j$ is given by $d_x(i,j)=x_i-x_j$ and $d_y(i,j)=y_i-y_j$, respectively, assuming a lattice constant of unity and respecting the periodic boundaries. The Pauli matrices $\sigma$ and $\tau $ operate in spin and particle-hole-space, respectively, and are connected by a tensor product. $N$ is the number of sites in the system. The summation over $<i,j>_m$ runs over all combinations up to the $m$-th nearest neighbor. In the above Hamiltonian $m=1$, i.e. only hopping between nearest neighbors is included.\\

\subsection{Ground State Calculations}

We consider the local magnetization as a free parameter, not limited by a priori assumptions about the magnetic ground state, and identify the energetically most favorable configuration of the magnetization $\textbf{m}_i$ for given tight-binding parameters $J^{tb}$, $\mu$, $\Delta$, $B_z$, and $\alpha$. 
Subsequently, we find the ground state in the framework of the Metropolis Monte-Carlo algorithm \cite{MC2, MC1}. 
Finding the magnetic ground state, i.e., the magnetic configuration that minimizes the total energy of all occupied states, with a Monte-Carlo procedure directly within the tight-binding calculations is computationally very expensive as the time required to calculate the total energy scales as $O(N^3)$ with the system size $N$. To address this problem, we use an approximative approach introduced in \cite{1D}.
First, we generate $8000$ random magnetic configurations by choosing each spin $\vec{m}_j$ independently from a random uniform distribution on the unit sphere.
Then, we calculate the total energy of the electronic system, i.e., the sum of all eigenvalues of the Hamiltonian in Eq.~\eqref{H_TB} below the Fermi energy $E_F$ using the \textit{Kwant} code \cite{kwant}. In the last step, we fit the obtained energies to a classical Heisenberg type Hamiltonian, using a least square method, obtaining the constants of the classical model. To take into account the itinerant nature of electrons in 2D systems we pay particular attention to multi-spin interactions on the order of $S^4$.
The employed classical Heisenberg Hamiltonian on a square lattice with periodic boundary conditions is given by
\begin{equation}\label{H_H}
\begin{split}
H_{\text{HB}}= & \sum_{<i,j>_5}\vec{s}_i^\top J_{i,j}^{cl} \vec{s}_j  - \sum_i \vec{B}\cdot\vec{s}_i\\
& + \sum_{i} C_{1}^1 \big([\vec{s}_i\cdot \vec{s}_{i+(1,0)}][ \vec{s}_{i+(1,1)} \cdot \vec{s}_{i+(0,1)}]\big) \\
& + \sum_{i} C_{2}^1 \big([\vec{s}_i\cdot \vec{s}_{i+(1,1)}][ \vec{s}_{i+(1,0)} \cdot \vec{s}_{i+(0,1)}]\big)\\
& + \sum_{i} C_{3}^1 \big([\vec{s}_i\cdot \vec{s}_{i+(0,1)}][ \vec{s}_{i+(1,0)} \cdot \vec{s}_{i+(1,1)}]\big)\\
& + \sum_{i} C_{1}^2 \big([\vec{s}_i\cdot \vec{s}_{i+(1,1)}][ \vec{s}_{i+(2,0)} \cdot \vec{s}_{i+(1,-1)}]\big) \\
& + \sum_{i} C_{2}^2 \big([\vec{s}_i\cdot \vec{s}_{i+(2,0)}][ \vec{s}_{i+(1,1)} \cdot \vec{s}_{i+(1,-1)}]\big)\\
& + \sum_{i} C_{3}^2 \big([\vec{s}_i\cdot \vec{s}_{i+(1,-1)}][ \vec{s}_{i+(1,1)} \cdot \vec{s}_{i+(2,0)}]\big), 
\end{split}
\end{equation}
where the indices $i$ and $j$ are vector valued containing the x-y-coordinate. $J_{i,j}^{cl}$ is a 3x3-matrix that includes all possible linear $2$-spin interactions.
The $4$-spin terms $C_k^n$ build a complete linear independent basis for all isotropic $4$-site-$4$-spin interactions \cite{tensors-iso}. We exclude otherwise considered $3$- and $2$-site-$4$-spin interactions for simplicity \cite{Laszlo}.
The interaction strengths $J_{i,j}^{cl}$, $C_{1}^n$, $C_{2}^n$, $C_{3}^n$ are translationally invariant, e.g., $J_{i+a,j+a}=J_{i,j}$, but not a priori directionally invariant. $\vec{B}$ is a vector that represents an external magnetic field and is constant over the whole lattice.
The spin interactions are visualized in Fig.~\ref{spin_interactions}. The summations for $2$-spin interactions run over all combinations of spins up to the 5th nearest neighbor, which are induced in multi-hopping processes of the tight-binding model. 
The $4$-spin interactions consider the two smallest possible squares of sites, i.e., $(0,0),(1,0),(1,1),(0,1)$ and $(0,0),(1,1),(2,0),(1,-1)$ and all translationally equivalent ones. The index $n$ denotes which of these two square types is chosen.
To find all $J_{i,j}^{cl}$, $C_k^n$, and $\vec{B}$, we fit the Heisenberg model to the energy dependence of the tight-binding model Eq.~\eqref{H_TB} using the Levenberg-Marquardt algorithm implemented in \textit{SciPy} \cite{scipy}, see Appendix A.
The resulting classical Heisenberg Hamiltonian is then used in a Metropolis Monte-Carlo simulation to determine the zero temperature ground state.
To generate the input for the fit, we use a 15x15 site system. Further increase of the system size does not lead to a significant change in the determined constants in Eq.~\eqref{H_H}.
As we are able to calculate the fitting input from a smaller system, and the change in energy from changing a singular spin in the Metropolis Monte-Carlo simulation can be calculated locally in a classical spin model, the required calculation time for each energy calculation does not scale with system size. Thus, this provides a very efficient method to find ground states of large systems.

\subsection{Heisenberg Parameters}

\begin{figure}
          \subfloat{\includegraphics[width=0.85\linewidth]{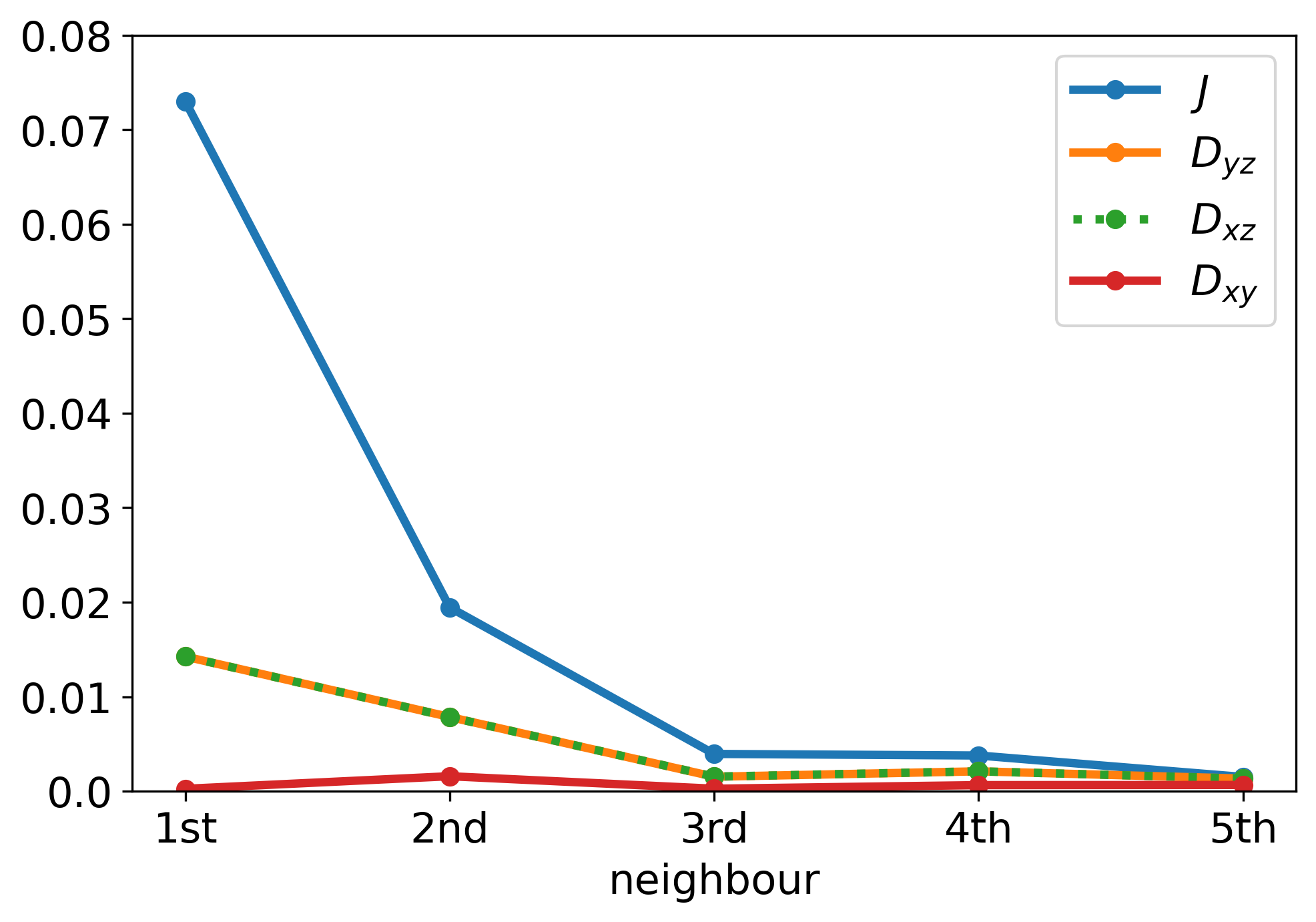}}   
\caption{Two-spin interactions. Absolute values of diagonal and off-diagonal elements of $J_{ij}^{cl}$ for $\alpha=0.2t$, $\Delta=0.5t$ and $B_z=0.0t$ with respect to the neighbors averaged over $J^{tb}$ and $\mu$, where the variance of the fit is lower than $0.2$.}
\label{Heisenberg_averages}
\end{figure}

\begin{figure}[h]
          \subfloat{\includegraphics[width=0.99\linewidth]{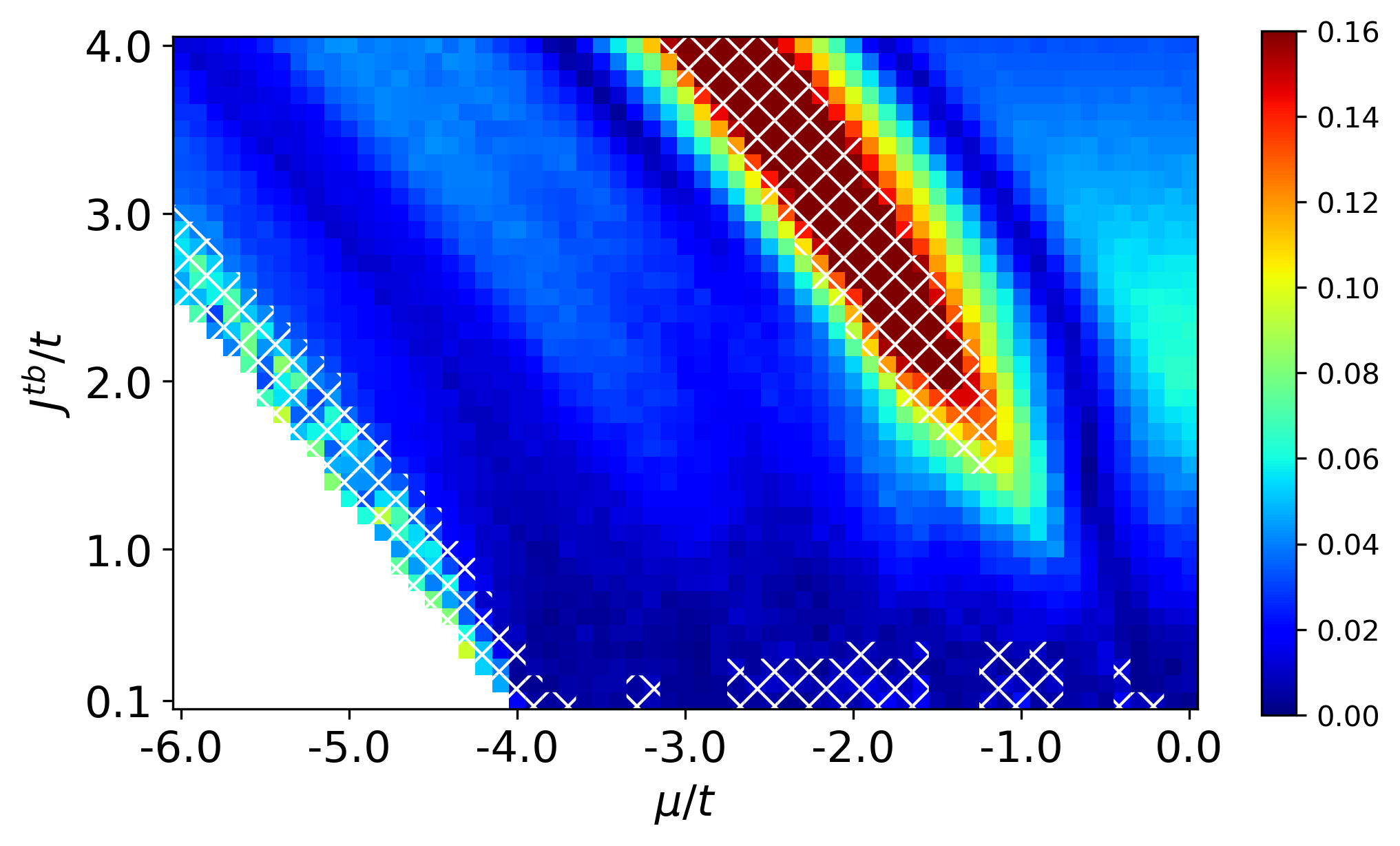}}
\caption{Relative strength of $4$-spin interactions for $\alpha=B_z=\Delta=0$ in dependence on $J^{tb}$ and $\mu$, calculated by the sum of the absolute values of all $C_l^n$ divided by the sum of the absolute values of all parameters in Eq.~\eqref{H_H}.
White denotes regions with largely unoccupied bands. In the hatched area, the variance of the fit exceeds 15\%, which we consider unreliable, see Appendix A.}
\label{spin4}
\end{figure}

In the following, we discuss our results of the fitted classical magnetic parameters.
First, we can describe the components of $J_{ij}^{cl}$ as exchange and Dzyaloshinskii–Moriya interaction (DMI).
\begin{equation}
\begin{split}\label{mapping}
    \vec{s_i} J_{ij}^{cl} \vec{s_j} & \approx 
    \vec{s_i}\begin{pmatrix}
        J & D_{xy} & -D_{xz} \\
        -D_{xy} & J & D_{yz} \\
        D_{xz} & -D_{yz} & J
    \end{pmatrix}\vec{s_j} \\
    & = J\vec{s_i}\cdot\vec{s_j} +
    \vec{D}\cdot \vec{s_i}\times\vec{s_j}
\end{split}
\end{equation}
with the classical exchange $J$ and the DMI-vector $\vec{D}=(D_{yz},D_{xz},D_{xy})^T$. The coupling constant $J$ and the components of $\vec{D}$ depend on the combination of sites $i$ and $j$, which is left out in Eq.~\eqref{mapping} for better readability. Fig.~\ref{Heisenberg_averages} shows the average of $J$ and $\vec{D}$ with respect to the distance between $i$ and $j$ averaged over all $J^{tb}$ and $\mu$ for $\alpha=0.2t$, $\Delta=0.5t$ and $B_z=0.0t$. Changing the superconducting order parameter $\Delta$ and the magnetic field $B$ within the considered parameter range only causes minimal changes to these averages. Setting the spin-orbit coupling $\alpha=0$ removes the off-diagonal elements and increases the strength of the diagonal elements.
In this case, we find that the diagonal elements of the matrices $J_{ij}^{cl}$ are always nearly identical along the diagonal, i.e., the variance of diagonal elements is around $0.01\%$, meaning that we find no anisotropy for the diagonal elements and they can be understood as the classical exchange interaction with a scalar product $\vec{s_i}\cdot\vec{s_j}$ as in Eq.~\eqref{mapping}, which is consistent with the symmetry of the initial tight-binding model. We also observe that the exchange interaction becomes significantly smaller on average with increasing distance, showing that long distance interactions are less important than short distance interaction, as one would expect, see Fig.~\ref{Heisenberg_averages} for representative parameters.\\
The off-diagonal elements can be mapped to DMI-vectors, which are mostly lying in the xy-plane, i.e. the $D_{xy}$ component is very small, and are oriented perpendicular to the vector between the respective sites. They obey the system's rotational and translational lattice symmetries.\\
We also find that $B_z$ in the tight-binding model directly translates to a $\vec{B}$-field in $z$-direction in the classical model, as expected.
In Fig.~\ref{spin4}, the relative strength of the $4$-spin interaction is shown, which is calculated as the sum of the absolute values of all $C$ divided by the sum of the absolute values of all fitted parameters in Eq.\eqref{H_H}. We found a small, but still significant contribution of $4$-spin interaction in well-fitted regions. This indicates that this simple seeming tight-binding system hosts complex multi-spin interactions beyond $2$-spin interactions.
The largest $4$-spin contribution is found in a region, where the fitting quality is not sufficient. Because the addition of $4$-spin interactions improves the fitting quality in that region, we speculate that this region is dominated by higher order or further reaching multi-spin interactions.
Adding non-vanishing RSO of $\alpha=0.2t$ causes only minimal changes in the pattern of $4$-spin interactions. The relative strength is lowered by $\sim 28\%$ on average, mostly due to an absolute increase in the strength of $2$-spin interactions due to the addition of DMI.

\subsection{Classification of Magnetic Phases using Contrastive Clustering} \label{classification}
\begin{figure}
    \centering
    \includegraphics[width=0.5\textwidth]{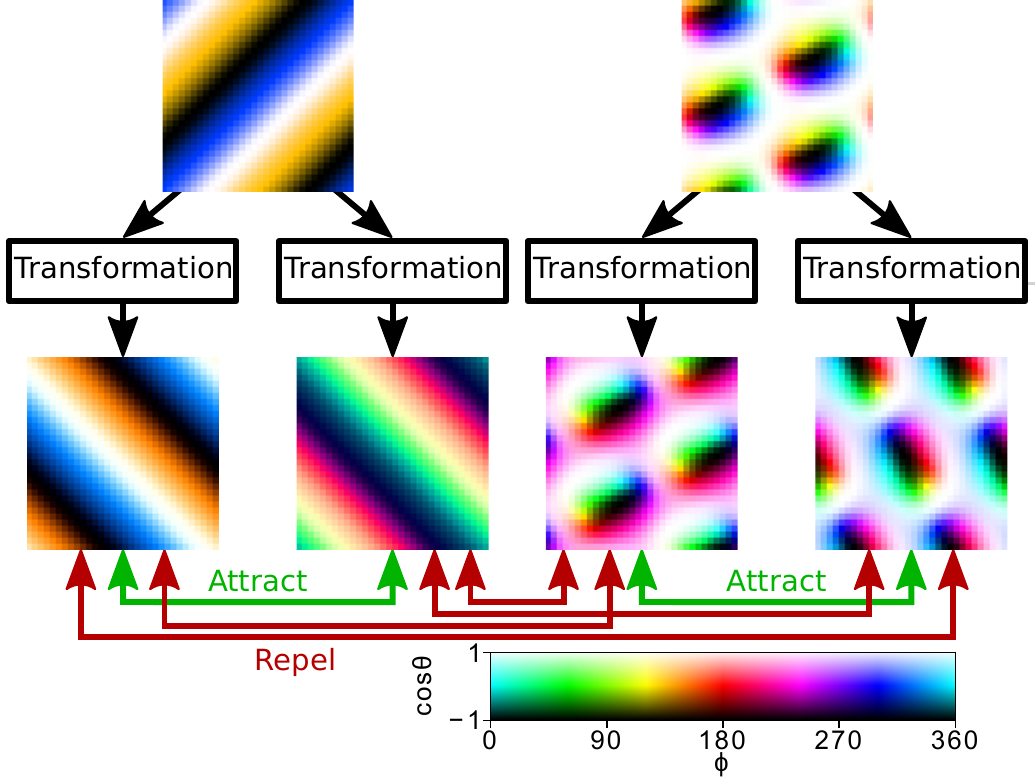}
    \caption{Visualization of the transformations used for the contrastive learning procedure: The top row shows two samples from the dataset, a spin spiral, and a skyrmionic configuration. We apply four different transformations to each sample: A random rotation in real space by 0°, 90°, 180°, or 270°, a random translation, and a random $\text{SO}(3)$ rotation of all spins. All transformations are chosen uniformly randomly and are equally likely. The objective of contrastive learning is to put transformed samples from the same original configuration close together in a latent space and repel samples that do not come from the same configurations.}
    \label{fig:cl_trans}
\end{figure}

\begin{figure}[t]
          \subfloat{\includegraphics[width=0.99\linewidth]{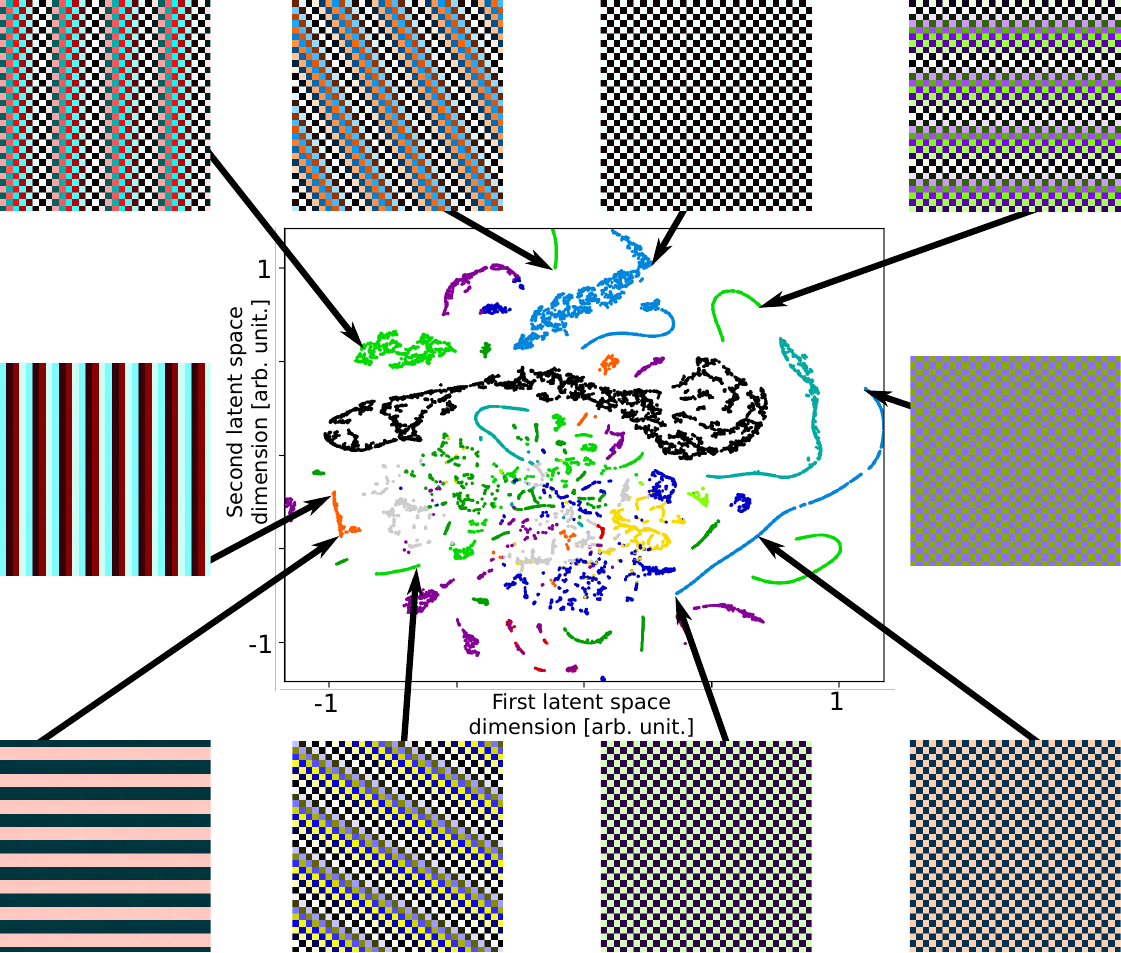}}
\caption{Similar magnetic configurations identified by Contrastive Clustering and their location in latent space. Each point in the scatter plot (center) corresponds to a configuration in the dataset. The 128-dimensional latent space is reduced to a two-dimensional space by t-distributed stochastic neighbor embedding (t-SNE)~\cite{van2008visualizing} for visualization. Each point is colored corresponding to the assigned phase with the colors being the same as in Fig.~\ref{phases_basic}. The color map for the configurations is the same as in Fig.~\ref{fig:cl_trans}. Arrows indicate the locations of ten representative configurations in the latent space. More samples can be viewed and explored online~\cite{LatentSpaceExplorer}.}
\label{samples_in_latentspace}
\end{figure}

To better understand the ground state of the Heisenberg model with $B$-field and $4$-spin interactions described in Eq.~\eqref{H_H}, we employ Contrastive Clustering~\cite{li2021contrastive} an unsupervised learning technique where the learning of a latent representation and cluster assignment are performed simultaneously by comparing different samples. The samples get encoded into points in a representation space, the latent space. The goal is to attract points in this space if they correspond to the same phase and repel them if they come from different phases. To achieve this, we apply symmetry transformations to magnetic configurations. Two transformed images attract each other if they originate from the same configuration, else they repel, as visualized in Fig.~\ref{fig:cl_trans}.
The details of the implementation of Contrastive Clustering follow Ref.~\cite{li2021contrastive}. We use Resnet18~\cite{he2016deep} as the encoder network, Adam~\cite{kingma2014adam} to train the network, and train our network for 2000 cycles over the whole dataset ensuring sufficient convergence of our loss function. The output of Resnet18 is a 512-dimensional vector. On top of that, two multilayer perceptrons are applied as projection heads, which project the output of the encoder to a 128-dimensional instance space (visualized in Fig.~\ref{samples_in_latentspace}) and a 40-dimensional clustering space. They both consist of one hidden layer with 512 neurons and a rectified linear unit (ReLU) as activation function. To deal with the unbalanced nature of our dataset, we adapted an oscillating computational temperature in the contrastive loss, as this has been shown to perform better with long-tail data~\cite{kukleva2023temperature}, i.e., data where a small part of classes have a large number of sample points but the others are associated
with only a few samples~\cite{zhang2023deep}. We implement the best-performing parameters of Ref.~\cite{kukleva2023temperature}.
For the transformations, we use intrinsic symmetries of the magnetic structure identification, which are a rotation of the whole lattice by either 0°, 90°, 180°, or 270°, a shift of the lattice by a random amount in the x and y directions, and rotation of all spins together in a random direction. This last step violates the axial anisotropy of the system when the Rashba-spin-orbit coupling $\alpha$ or the magnetic field $\vec{B}$ is non-zero. We successfully classify the unlabeled spin data into different phases. Hence, we create a phase diagram without having to specify the phases in advance, which allows the detection of unexpected phases.
The latent space after the training procedure can be seen in Fig.~\ref{samples_in_latentspace}. We provide an online tool for a more in-depth exploration of the latent space and the data~\cite{LatentSpaceExplorer}.

We categorize the magnetic phases as follows.
First, the artificial neural network splits the magnetic ground states into $40$ clusters. Often, multiple clusters belong to the same magnetic phase, e.g., there are multiple clusters with harmonic spin spirals but different ranges of relative angles between the nearest magnetic moments. Those clusters are then manually combined to $11$ magnetic phases, which we identify by spot checks. The number of clusters is chosen larger than the number of expected physical phases to avoid misclassification. For example, when limited by too few clusters, the ANN could assign an $87^\circ$-spiral to an antiferromagnetic cluster instead of a spiral one. When a cluster cannot be clearly assigned to a single magnetic phase, i.e., we identify different magnetic phases within one cluster, we label that cluster as \textit{mixed}.
Three clusters initially labeled as \textit{mixed} could be separated by adding an additional condition. For one of those, all states with $\mu<-2.0t$ belong to a magnetic 2x2-pattern, while the rest belong to AFM spin spirals. The other two contain harmonic spin spirals and 2q-spin-spirals but for $\alpha=0.0$ only contain harmonic spin spirals.
Additionally, the artificial neural network (ANN) often combines skyrmion lattices and spin spirals with a periodicity that is similar to the skyrmion size into one cluster.
To identify skyrmionic phases, we calculate the skyrmion number $N_{sk}$, see Eq.~\eqref{Nsk} and assign everything with $|N_{Sk}|>0.95$ as part of a skyrmionic phase, giving some tolerance for numerical precision.

\section{Results} \label{results}
For the sake of clarity, we limit ourselves to two distinct values for each of the parameters $\Delta$, $B_z$ and $\alpha$, representing the presence or absence of the corresponding physical properties, while scanning for a wide range of $J^{tb}$ and $\mu$. We start by providing an overview of the kinds of magnetic ground states found in our calculations. 
Then, we describe systems with vanishing Rashba-spin-orbit coupling and vanishing magnetic fields, with and without non-vanishing superconductivity. Afterwards we show the influence of non-vanishing RSO in combination with an external magnetic field, which typically stabilizes skyrmions.

\begin{figure*}[t]
          \subfloat{\includegraphics[width=0.99\linewidth]{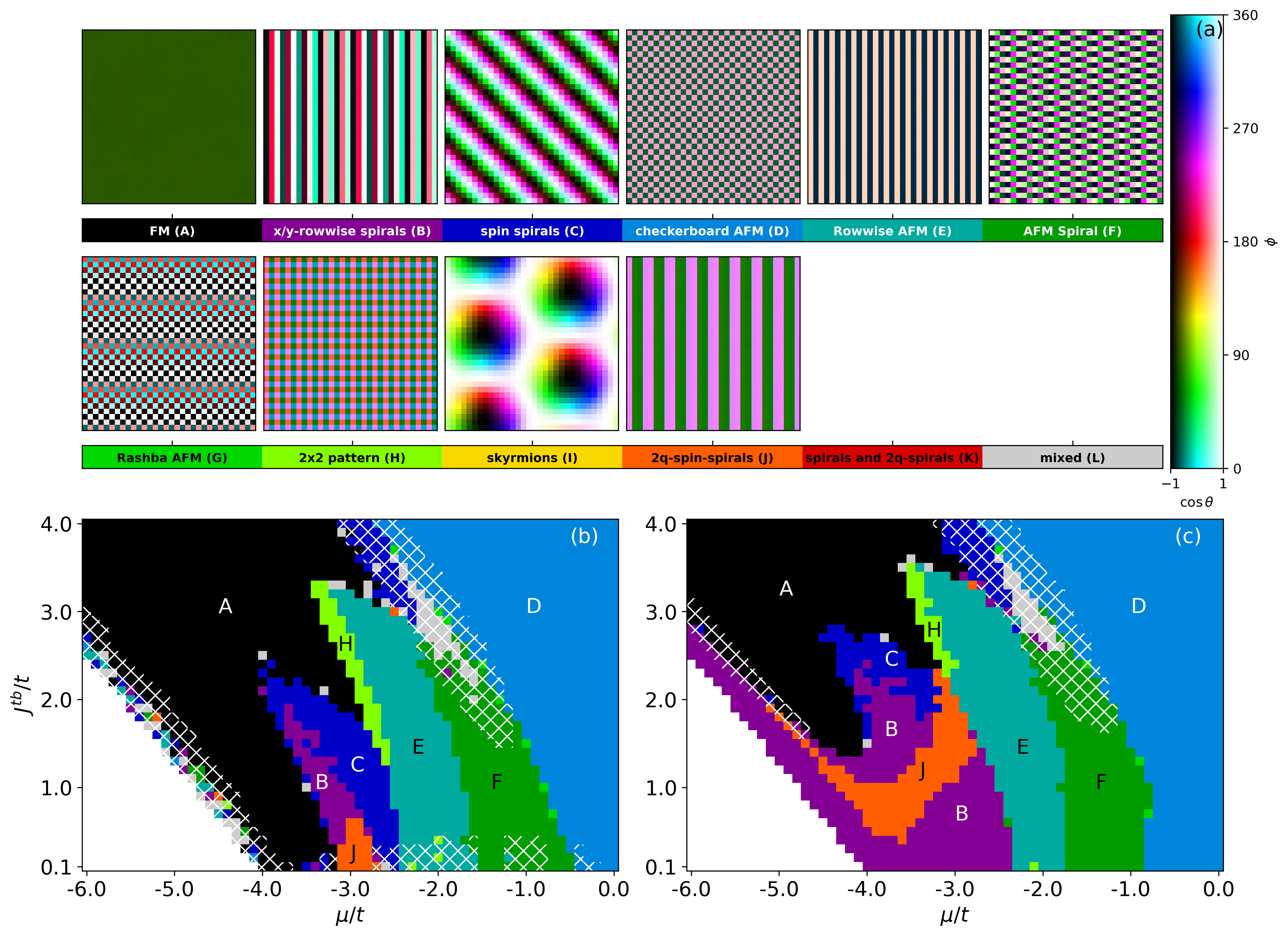}}
\caption{Magnetic ground states for $\alpha=0$, $B_z=0$ and (b) $\Delta=0.0$, (c) $\Delta=0.5 t$ with respect to $\mu$ and $J^{tb}$. The color white denotes regions with largely unoccupied bands. The hatched area leaves the validity regime of the model (see Fig. \ref{spin4}).
 The colorbar for (b) and (c) is given in (a) with examples for each type of magnetic ground state, where the color denotes the orientation of the spins with each pixel representing one spin. For "spin spirals and 2q-spin-spirals" and "mixed" no examples are given.}
\label{phases_basic}
\end{figure*}

\subsection{Vanishing Rashba-Spin-Orbit Coupling}

\begin{figure*}[t]
          \subfloat{\includegraphics[width=1.0\linewidth]{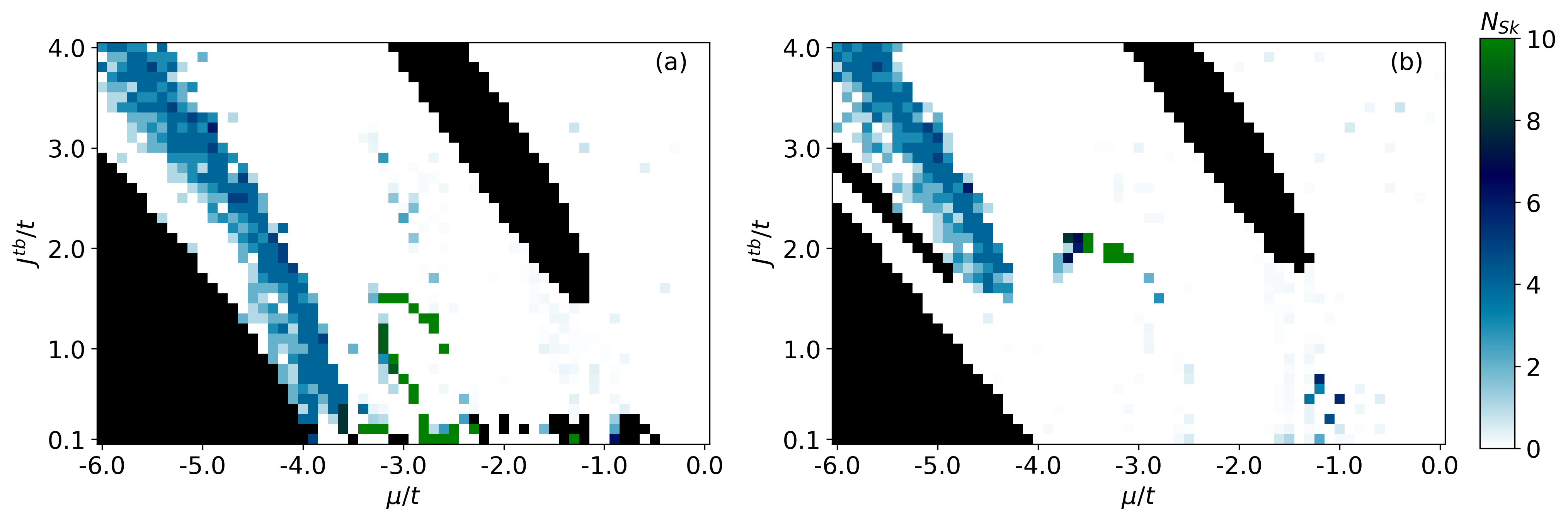}}                 
\caption{Absolute value of the skyrmion number $|N_{sk}|$ (denoted by the color) with respect to $J^{tb}$ and $\mu$ for $\alpha=0.2t$, $B_z=0.02t$ and (a) $\Delta=0.0$, (b) $\Delta=0.5 t$. The blacked out regions leaves the validity regime of the model and is equivalent to the hatched area in Fig. \ref{spin4}.}
\label{skyrmions}
\end{figure*}

We calculate the magnetic ground states of a $32\times 32$ system with periodic boundary conditions in real space. The ANN-classified magnetic state as described in Sec.~\ref{classification}, in dependence on the magnetic coupling $J^{tb}$ and the chemical potential $\mu$ are shown in Fig.~\ref{phases_basic} for $\Delta=0.0$ and $\Delta=0.5t$ without RSO, i.e., $\alpha=0$.
Fig.~\ref{phases_basic} b and c show phase diagrams for $\Delta=0.0$ and $\Delta=0.5t$, respectively. Examples of ground states are presented in Fig.~\ref{phases_basic} a.

We first provide an overview over the types of identified magnetic ground states.
The \textit{ferromagnetic} (FM) state describes a state in which all spins are parallel.
In the case of \textit{x/y-row-wise spirals} and \textit{spin spirals} the spins are aligned parallel in one direction and as a harmonic spin spiral in the perpendicular direction, i.e., the relative angle between spins remains constant in the propagation direction of the harmonic spin spiral. In the case of the \textit{x/y-row-wise spin spirals}, the propagation direction of the spin spiral is either the x- or y-axis, while the other \textit{spin spirals} summarize all other spiral directions and clusters, that were mixed between the x/y direction and other directions.
The \textit{checkerboard AFM} phase is defined by all spins being aligned anti-parallelly to all of their nearest neighbors. 
The \textit{row-wise AFM} has parallel spins along one direction and anti-parallel nearest neighbors along the perpendicular direction. 
\textit{AFM spin spirals} correspond to the spins being aligned as harmonic spin spirals in one direction and anti-parallel in the perpendicular direction. 
The \textit{Rashba AFM} are specific AFM spin spirals that can be described as a superposition of a \textit{checkerboard AFM} and a harmonic spin spiral with an angle that is characteristic for the chosen RSO strength $\alpha$, i.e., constant for a chosen $\alpha$.
We observe no inherent preference for the orientation within the x-y-plane for the \textit{Rashba AFM}, \textit{row-wise AFM} and \textit{AFM spin spiral}.
We also identified a phase with a \textit{2x2 pattern}, where we find a repeated structure of 2 times 2 spins with the following two additional properties. First, along the x- and the y-axis the relative angles are constant in their absolute value but alternating in sign along each axis, e.g., the relative angle along the x-axis alternates between $+80^\circ$ and $-80^\circ$. Second, all spins are anti-parallel to their next-nearest, i.e., diagonal, neighbor.
We labeled a phase as \textit{skyrmions}, if we find a skyrmion number $|N_{sk}|>0.95$, which is calculated by
\begin{equation}\label{Nsk}
    N_{Sk}=\frac{1}{4\pi} \sum_\text{sites} \vec{M}\cdot\left(\frac{\partial \vec{M}}{\partial x}\times \frac{\partial \vec{M}}{\partial y}\right) ,
\end{equation}
where $\vec{M}$ is the local magnetization and where we calculate the derivatives of the continuum extrapolation by the inverse Discrete Fourier Transform (DFT) of the product of the DFT frequencies and the DFT of the spins following $f'(x)=\frac{1}{N}\sum_k e^{\mathrm{i} k x}(\mathrm{i} k)\sum_z e^{-\mathrm{i} k z} f(z)$. This avoids numerical artifacts for antiferromagnetic states that can appear with lattice adapted invariants \cite{bergDefinitionStatisticalDistributions1981} for the cost of deviations from the integer quantization of the skyrmion number. In most skyrmionic cases, we observe a skyrmion lattice, while in some cases singular skyrmions appear in a ferromagnetic background. In \textit{2q-spin-spirals}, we observe parallel spins in one direction and following a pattern of two alternating relative angles $\theta_1$ and $\theta_2$ in the perpendicular direction. For $\alpha=0$, these two angles average to $(\theta_1 + \theta_2) /2=\pi/2\pm 0.01$ in $85\%$ of cases. While in the harmonic FM and AFM spin spirals magnetization rotates with a wave-vector q along high symmetry crystallographic directions, the multi-q-spin-structures are superpositions of several rotations with high symmetry q-vectors \cite{2q}.
\textit{Spin spirals and 2q-spin-spirals} denote that the ANN cannot differentiate between 2q-spin-spirals and harmonic spin spirals, which only occurs for $\alpha=0.2$.
\textit{Mixed} denotes that the ANN created a label which we could not assign a definite magnetic state to.

Before discussing the influence of superconductivity, we start with $\Delta=0$, i.e., vanishing superconductivity. For large $J^{tb}>5.0t$, we only observe ferromagnetic (FM) and antiferromagnetic (AFM) phases. For smaller $J^{tb}$, we observe rich magnetic phases between the FM (A) and AFM (D) phase shown in Fig.~\ref{phases_basic}b.
For chemical potentials $-1.7t<\mu<-0.9t$,  we find AFM spirals (F). At $\mu<-1.7t$, this phase transitions into row-wise AFMs (E). For lower chemical potentials, we find harmonic spin spirals with varying propagation directions other than the x- or y-axis (C). Lowering the chemical potential further results in x-y-row-wise spin spirals (B) and then in an FM phase. For $J^{tb}>1.2t$, the row-wise AFM phase first transitions into a 2x2 pattern (H) and then into the FM phase. Additionally around $\mu=-3.0t$ and $J^{tb}<0.4t$, there is a small region in parameter space with 2q-spin-spirals (J).

For non-vanishing superconductivity, $\Delta=0.5t$, the position and the shape of the magnetic phases change slightly. The AFM (D) and AFM spiral (F) phases remain mostly unchanged.
The area of the 2x2 pattern (H) becomes slightly smaller, being replaced by row-wise AFMs (E) for $J^{tb}<2.2t$.
The x/y-row-wise spin spiral phase (B) grows considerably and within it, the 2q-spin-spiral phase (H) moves towards larger $J^{tb}$ and also expands. As large parts of the latter follow an $\up\up\down\down$ pattern, this stands in contrast to results from the same Hamiltonian in 1D \cite{1D}, where increased superconductivity causes this phase to vanish.
The spin spiral phase (C) moves towards larger J remaining at the transition between the FM (A) and the x/y-row-wise spin spirals (B).

\subsection{Non-Vanishing Rashba-Spin-Orbit Coupling}

\begin{figure*}[t]
          \subfloat{\includegraphics[width=0.99\linewidth]{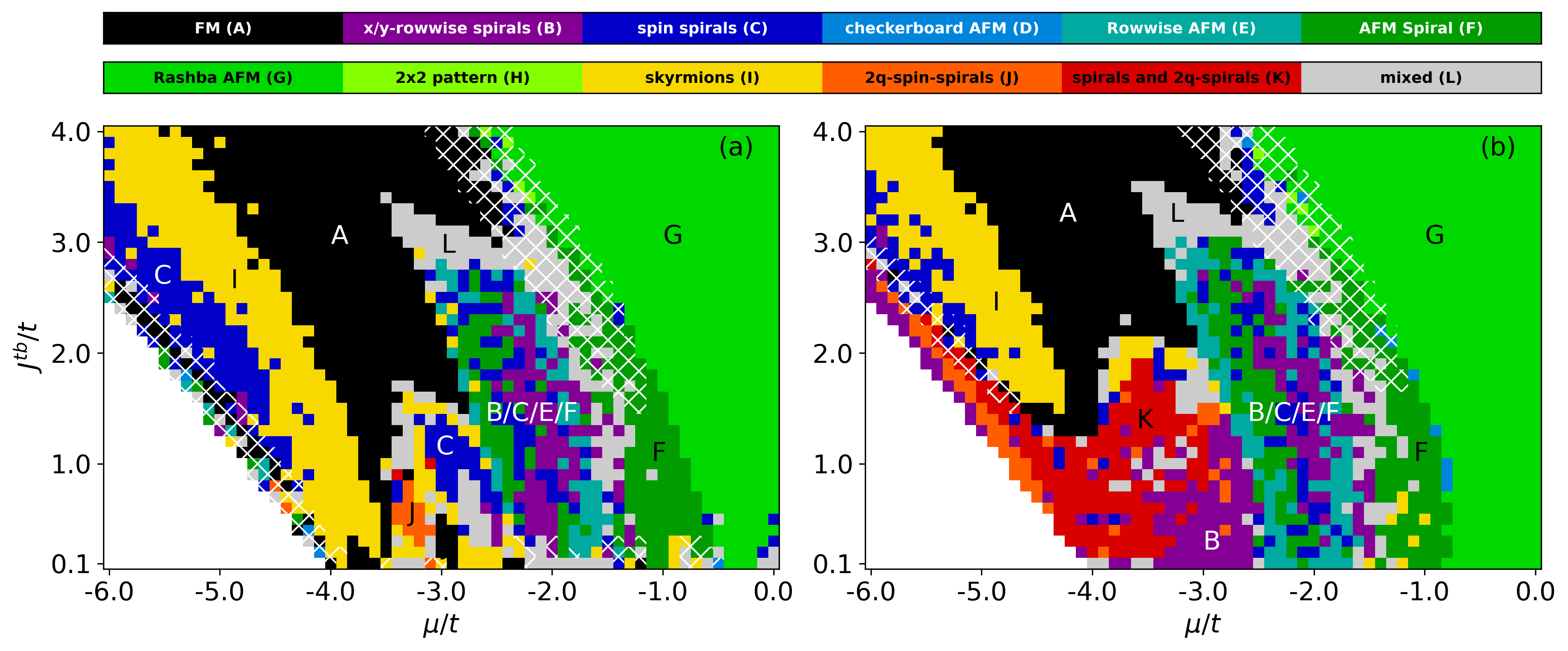}}
\caption{Magnetic ground states in systems with Rashba spin-orbit coupling $\alpha = 0.2t$, an additional out-of-plane magnetic field $B_z = 0.02t$ and (a) $\Delta=0.0$, (b) $\Delta=0.5 t$ with respect to $\mu$ and $J^{tb}$. The color white denotes regions with largely unoccupied bands. The hatched area leaves the validity regime of the model (see Fig. \ref{spin4}).}
\label{RSO}
\end{figure*}

Skyrmion lattices can be stabilized by an external magnetic field perpendicular to the lattice \cite{stableSkyrmions,stableSkyrmion2,stableSkyrmion3}, 
We therefore added RSO $\alpha=0.2t$ and a small magnetic field $B_z=0.02t$ and succeeded in finding a parameter region with skyrmion lattices. The magnetic field breaks the rotational symmetry of the Hamiltonian.
Fig.~\ref{skyrmions} shows the skyrmion number with respect to the magnetic coupling strength $J^{tb}$ and the chemical potential $\mu$ for superconducting order parameters $\Delta=0$ and $\Delta=0.5t$. An extended large skyrmionic phase emerges, which gradually moves to larger $J^{tb}$, when the superconducting order parameter $\Delta$ is increased.\\
The effect of RSO and a magnetic field on the other phases is displayed in Fig.~\ref{RSO}.
Besides the appearance of a new skyrmionic and spin spiral phase for low $\mu$ within the FM phase, the overall shape of the phases remains largely unchanged, but the magnetic structures within the phases change.
The previous checkerboard AFM phase becomes a checkerboard AFM in superposition with a spin spiral of a characteristic angle of $0.124\pi$, which we called \textit{Rashba AFM} here.
Most parts of the FM phase remain ferromagnetic, but towards lower $\mu$, an additional skyrmionic and spin spiral phase occurs.
The inner parts of the row-wise AFM phase is replaced by spin spirals, of which only some are AFM spirals.
The 2x2 pattern does not occur anymore.
For $\Delta=0.5t$, the 2q-spin-spiral phase still exists, but the ANN fails to differentiate it from harmonic row-wise spin spirals.
We also observe a larger area in the phase diagram labelled B/C/E/F, where multiple phases coexist. When inspecting single samples in this area, we find that the phases are accurately assigned by the ANN and that this apparent disorder truly reflects the large diversity of magnetic configurations in that area.

\section{Conclusions} \label{conclusions}
Our results show that a simple effective model for magnetic moments coupled to a superconducting substrate with Rashba spin-orbit coupling accommodates an unexpectedly rich phase diagram, helping with the search for exotic and noncollinear phases at magnetic surfaces, which can be effectively categorized by an ANN with contrastive learning algorithm. The combination of the 2D tight-binding model with Monte Carlo simulations permits us to comprehensively categorize the magnetic ground states in the SC and non-SC regimes. In addition to the collinear FM and AFM configurations as well as harmonic noncollinear states known from one-dimensional models, 2D systems accommodate manifold multi-q noncollinear structures and non-trivial collinear configurations like Rashba-AFM or 2q-spin-spirals. An interesting property of an ultrathin film on a square lattice is the appearance of multi-spin interactions that have been predicted to exist in itinerant magnets but are only important in few experimental studies so-far like the nanoskyrmion lattice in Fe/Ir(111) \cite{Heinze2011}.
Our phase diagrams help researchers working with ultrathin magnetic films, showing what types of magnetic ordering, they can expect in their system or, vice versa, to estimate model parameters from the observed magnetic ordering. Generally, superconductivity decreases the area of the FM phase in favor of AFM and non-collinear structures, while the Rashba coupling favors noncollinearity.\\
Given that chiral structures are promising candidates to induce non-trivial electronic topological states, one could further explore the electronic properties of the presented magnetic systems with the described self-consistent model. This includes skyrmionic, multi-Q, and spiral magnetism in combination with a non-vanishing superconducting order parameter and potentially topological insulators \cite{Baum2015,Bedow2020_3Q,Kubetzka2020,mascot2021topological}. This finding prompts to look further into the electronic topological properties of the described system in external magnetic fields and for different anisotropy types. Specifically, an investigation of the electronic topology of a magnetic island with open boundary conditions instead of periodic boundaries could reveal whether skyrmions and Majorana modes can occur naturally in the same sample without a preset magnetization \cite{mascot2021topological}. The presented method of finding magnetic ground states of tight-binding models can also be applied to more complex systems. One possible direction is to create a model of a complex artificial array of magnetic atoms on a superconducting surface \cite{chain2d}, and find the dependence of magnetic ground states on the symmetries of this array using the proposed method of reducing to a model containing only the magnetic moments, which is especially efficient when only a small fraction of the total system is magnetic.

\section*{Acknowledgments}
The authors thank Eric Mascot for helpful discussions. J.N., T.P., and R.W. gratefully acknowledge financial support of the Cluster of Excellence 'Advanced Imaging of Matter' (EXC 2056 - project ID 390715994) of the Deutsche Forschungsgemeinschaft (DFG) and from the European Union via the ERC Advanced
Grant ADMIRE (project No. 786020). 
T.P. acknowledges funding by the DFG (project no. 420120155) and the European Union (ERC, QUANTWIST, project number 101039098).
E.Y.V. and T.M. acknowledge financial support provided by the Deutsche
Forschungsgemeinschaft (DFG) via Project No. 514141286.
\newpage

\section*{Appendix A: Discussion on the fitting quality and reliability of the magnetic model}
\begin{figure}[h]
          \subfloat{\includegraphics[width=0.99\linewidth]{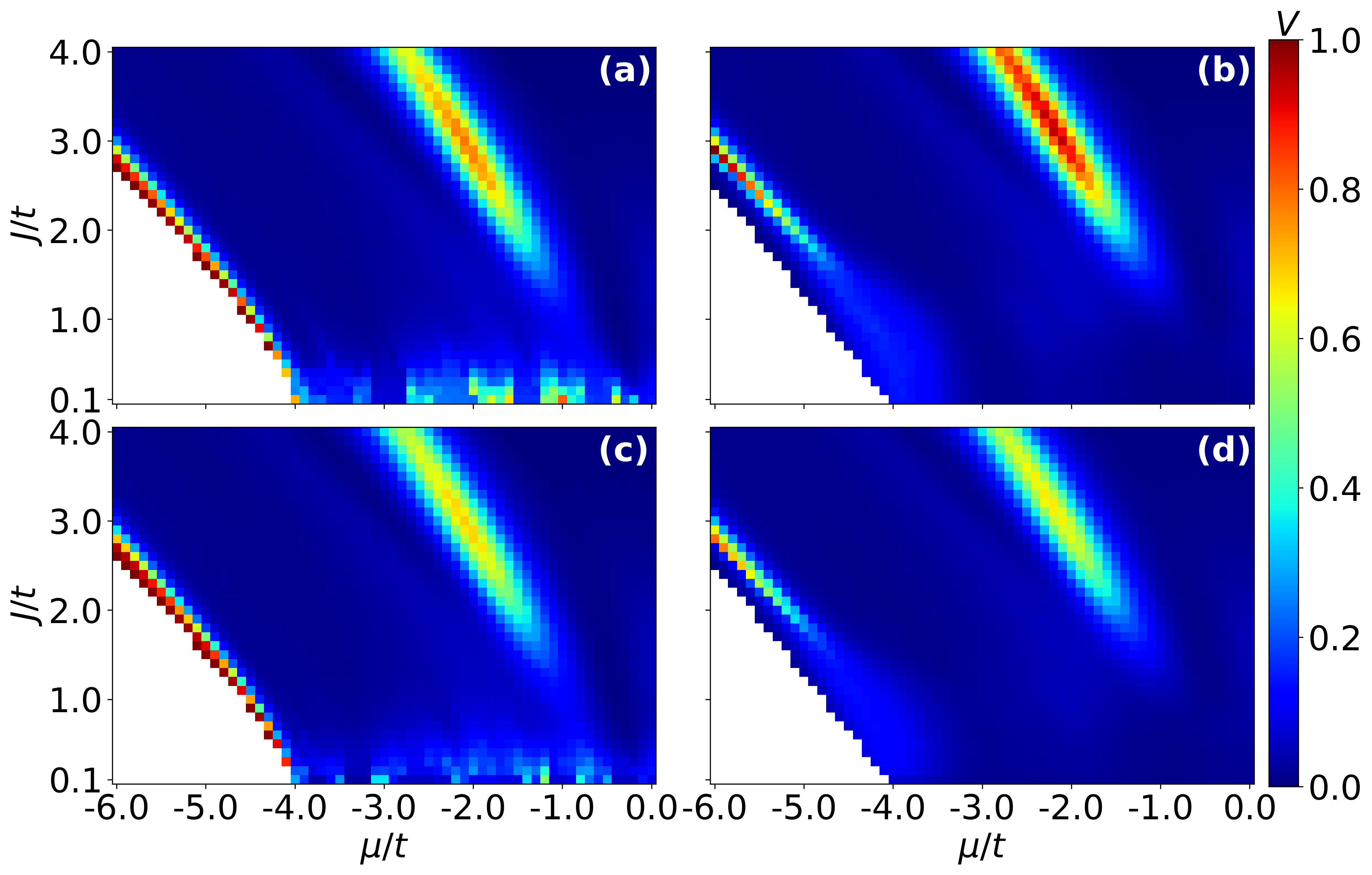}}
\caption{Normalized variance as a measure for fitting quality for (a) $\alpha=0, B_z=0, \Delta=0$, (b) $\alpha=0, B_z=0, \Delta=0.5t$, (c) $\alpha=0.2t$, $B_z=0.02t, \Delta=0$ and (d) $\alpha=0.2t, B_z=0.02t, \Delta=0.5t$ in dependence on $J^{tb}$ and $\mu$.
White denotes regions with largely unoccupied bands.}
\label{var}
\end{figure}
To judge the quality of the fits, we employ the normalized variance
\begin{equation}
    V=\frac{var(\vec{E}_{cl}-\vec{E}_{tb})}{var(\vec{E}_{tb})},
\end{equation}
where $\vec{E}_{cl}$ and $\vec{E}_{tb}$ contain the total energies calculated with the fitted classical model and the original tight-binding model, respectively.
The resulting variance is shown in Fig.~\ref{var} for the datasets used in the main text. For simultaneously small $\mu$ and $J$, no bands reach below the Fermi energy for $\Delta=0$, and for $\Delta>0$ the spin structure has no significant contribution to the total energy. At the edge of this region, bands are only occupied for some magnetic configurations, which results in bad fits.
The variance has a large peak next to the AFM region towards lower $\mu$. In this region the fitting quality is not good enough to make reliable assessments. However, we did observe that the fitting quality significantly increased (lowering the peak variance from 0.96 to 0.78) by including $4$-spin-interactions, while adding more long range interactions, i.e. 6th to 10th neighbor interactions, did not lead to any improvements. This suggests that higher-order interactions are required to fully understand the magnetism in this region. This is also in line with results from a similar model in 1D, where a parameter region left of the AFM phase could only be fitted by including $4$-spin-interactions \cite{1D}.\\
We also observe a high variance for small values of $J$, when $\Delta=0$.\\
In the main manuscript we crossed out regions, where the variance is larger than $0.2$ to display that the results in these regions are not reliable. Similarly, data close to the crossed out regions should also considered with caution, as the variance changes smoothly with the parameters of the tight-binding model.\\

We settled for a sample size of $N_{sample}=8000$. For larger sample sizes the variance does not increase further, showing that 8000 is sufficient to prevent overfitting.\\

\bibliography{BibFile}

\end{document}